\documentclass[amsmath,amssymb]{revtex4}
\usepackage{graphicx}
\usepackage{amscd,amsmath,amsthm,amsfonts,amssymb}
\def\[{\begin{equation}}
\def\]{\end{equation}}

\begin{document}
\title{Rogue Waves in (2+1)-Dimensional Three-Wave Resonant Interactions}
\author{
Bo Yang$^{1}$, Jianke Yang$^{2}$}
\address{$^{1}$ School of Mathematics and Statistics, Ningbo University, Ningbo 315211, China\\
$^{2}$ Department of Mathematics and Statistics, University of Vermont, Burlington, VT 05401, U.S.A}

\begin{abstract}
Rogue waves in (2+1)-dimensional three-wave resonant interactions are studied. General rogue waves arising from a constant background, from a lump-soliton background and from a dark-soliton background have been derived, and their dynamics illustrated. For rogue waves arising from a constant background, fundamental rogue waves are line-shaped, and multi-rogue waves exhibit multiple intersecting lines. Higher-order rogue waves could also be line-shaped, but they exhibit multiple parallel lines. For rogue waves arising from a lump-soliton background, they could exhibit distinctive patterns due to their interaction with the lump soliton. For rogue waves arising from a dark-soliton background, their intensity pattern could feature half-line shapes or lump shapes, which are very novel.
\end{abstract}
\maketitle

\section{Introduction}
Three-wave interaction arises in a wide variety of physical systems, such as water waves, nonlinear optics, plasma physics, and others \cite{Bloembergen1965,Benney_Newell,Kaupreview1979,Ablowitz_book,3wave_wateropt1,3wave_wateropt2,3wave_wateropt3,3wave_wateropt4,3wave_wateropt5}. When the wavenumbers and frequencies of the three waves form a resonant triad, this interaction is the strongest. In this case, the governing equations for this interaction are integrable \cite{Zakharov1973,Ablowitz_Haberman1975,Zakharov1975,Zakharov1976,Kaup1976,Kaup1980,Kaup1981,Kaup1981b}. As a consequence, multi-solitons in one spatial dimension and multi-lumps in two spatial dimensions of this system have been derived \cite{Zakharov1975,Zakharov1976,Kaup1976,Craik1978,Ablowitz_book,Zakharov_book,Kaup1981b,Gilson1998,Yang_Stud}.

Rogue waves are an interesting class of solutions in nonlinear wave systems that ``come from nowhere and disappear with no trace" \cite{Akhmediev_2009}. These waves have been linked to freak waves in the ocean and extreme events in optics \cite{Ocean_rogue_review,Pelinovsky_book,Solli_Nature,Wabnitz_book}. In addition, rogue waves have been observed in water tanks, optical fibers and plasma \cite{Tank1,Tank2,Tank3,Tank4,Plasma,Fiber1,Fiber2,Fiber3}. Due to their mathematical novelty and physical connections, rogue waves have attracted a lot of studies, especially in the integrable systems community, because
analytical expressions of rogue waves can be explicitly obtained when the nonlinear wave system is integrable. So far, rogue waves have been derived in a large number of integrable systems, such as the nonlinear Schr\"odinger (NLS) equation \cite{Peregrine,AAS2009,DGKM2010,ACA2010,KAAN2011,GLML2012,OhtaJY2012,DPMVB2013}, the derivative NLS equations \cite{KN_rogue_2011,KN_rogue_2013,CCL_rogue_Chow_Grimshaw2014,CCL_rogue_2017,YangYangDNLS}, the Manakov system \cite{BDCW2012,ManakovDark,LingGuoZhaoCNLS2014,Chen_Shihua2015,ZhaoGuoLingCNLS2016,YangYang2021b}, the Davey-Stewartson equations \cite{OhtaJKY2012,OhtaJKY2013}, and many others \cite{AANJM2010,OhtaJKY2014,ASAN2010,Chow,MuQin2016,ClarksonDowie2017,YangYangBoussi,JCChen2018LS,XiaoeYong2018}.
In addition, rogue waves that arise from a non-uniform background have also been reported in several wave systems \cite{Qin2015,Peli2019,He2021}.
These explicit solutions of rogue waves shed much light on the rogue dynamics in physical systems that are governed by the underlying integrable equations.

Rogue waves in (1+1)-dimensional three-wave resonant interactions have been derived and analyzed as well
\cite{BaroDegas2013,DegasLomba2013,ChenSCrespo2015,WangXChenY2015,ZhangYanWen2018,YangYang3wave}. Compared to rogue waves in many other (1+1)-dimensional equations, rogue waves in this three-wave system exhibited a much wider variety of solutions and wave patterns. This makes us wonder, what kind of rogue wave behaviors will arise in the (2+1)-dimensional three-wave system, i.e., the three-wave system with two spatial dimensions instead of one.

In this article, we study general rogue wave solutions in the (2+1)-dimensional three-wave system by the bilinear method.
Rogue waves arising from both a constant background and a non-constant background are derived, and our solutions are expressed through Schur polynomials. We find that for rogue waves arising from a constant background, fundamental rogue waves are line-shaped, and multi-rogue waves display multiple intersecting lines. Higher-order rogue waves could also be line-shaped, but would exhibit multiple parallel lines instead of intersecting lines. For rogue waves arising from a non-constant background, we report two different types of solutions. One type is rogue waves that emerge from a lump-soliton background, and these waves are rational solutions. The other type is rogue waves that emerge from a dark-soliton background, and these waves are semi-rational solutions which contain both rational and exponential components. We find that for the latter type of rogue waves, their intensity can exhibit half-line shapes or lump shapes, which are fascinating.

This paper is structured as follows. In Sec.~2, we introduce the (2+1)-dimensional three-wave system and their constant-background solutions. In Sec.~3, we present our rogue wave solutions that arise from a constant background. In Sec.~4, we illustrate dynamics of these rogue waves emerging from a constant background. In Sec.~5, we present our rogue wave solutions that arise from a non-constant background, including from a lump-soliton background and a dark-soliton background, and illustrate their dynamics. In Sec.~6, we provide derivations of rogue waves presented in Secs.~3 and 5. Sec.~7 summarizes the paper.

\section{Preliminaries} \label{sec:pre}
The (2+1)-dimensional three-wave resonant interaction system is
\begin{eqnarray}
&& \left(\partial_{t} + \mathbf{V}_{1} \cdot \nabla \right) q_{1}(x, y, t) = \epsilon_{1} q_{2}^* (x, y, t) q_{3}^*(x, y, t), \nonumber \\
&&  \left(\partial_{t} + \mathbf{V}_{2} \cdot \nabla \right) q_{2}(x, y, t)= \epsilon_{2} q_{1}^* (x,y, t) q_{3}^*(x,y, t),  \label{3WRIModel}\\
&&    \left(\partial_{t} + \mathbf{V}_{3} \cdot \nabla  \right) q_{3}(x, y, t) = \epsilon_{3} q_{1}^* (x,y, t) q_{2}^*(x,y, t). \nonumber
\end{eqnarray}
Here, $\nabla=\left( \partial_x, \ \partial_y \right)$ is the gradient operator in the $\mathbb{R}^2$ space, $\mathbf{V}_{k}=\left( V_{k,1}, V_{k,2} \right)$ are the three waves' group velocities along the $(x, y)$ directions, and the asterisk `*' represents complex conjugation. Adopting a coordinate system that moves at the velocity of the third wave, we make $\mathbf{V}_{3}=(0,0)$ without loss of generality. In addition, we assume that $\mathbf{V}_{1}$ and $\mathbf{V}_{2}$ are not parallel to each other, i.e., $V_{11}V_{22}-V_{12}V_{21}\ne 0$. Parameters $\epsilon_{j}$ are nonlinear coefficients that can be scaled to $\pm 1$. In addition, one can fix $ \epsilon_{1}=1$.

The above three-wave system admits plane-wave solutions
\begin{eqnarray}
&& q_{1,0}(x, y, t)= \rho_{1}  e^{\textrm{i} (k_{1}x + \lambda_{1}y + \omega_{1} t)},  \nonumber  \\
&& q_{2,0}(x, y, t)= \rho_{2}  e^{\textrm{i} (k_{2}x + \lambda_{2}y +\omega_{2} t)},   \label{PlanewaveSolu2}\\
&& q_{3,0}(x, y, t)= \textrm{i} \rho_{3}  e^{\textrm{i} (k_{3}x + \lambda_{3}y +\omega_{3} t)},   \nonumber
\end{eqnarray}
where wave vectors $(k_j, \lambda_j)$ and frequencies $\omega_j$ satisfy the resonance relations
\[
k_{1}+ k_{2}+ k_{3}=0, \quad \lambda_{1}+ \lambda_{2}+ \lambda_{3}=0, \quad \omega_{1}+ \omega_{2}+ \omega_{3}=0,
\]
and wave-amplitude parameters $(\rho_1, \rho_2, \rho_3)$ satisfy the following conditions
\begin{eqnarray}
&&  \rho_{1} \left( \omega_{1} + V_{1,1}k_{1}+ V_{1,2}  \lambda_{1} \right) = -\epsilon_{1} \rho_{2}^* \rho_{3}^*,  \nonumber \\
&&  \rho_{2} \left( \omega_{2} + V_{2,1}k_{2}+ V_{2,2}  \lambda_{2} \right) = -\epsilon_{2} \rho_{1}^* \rho_{3}^*,  \label{Pararlation2}\\
&& \rho_{3} \left( \omega_{1} +\omega_{2} \right) = \epsilon_{3} \rho_{1}^* \rho_{2}^*.  \nonumber
\end{eqnarray}
These plane waves have constant amplitudes in the $(x,y)$ space; so we will call them constant-background solutions.

In this article, we assume that the three wave amplitudes $|\rho_1|, |\rho_2|$ and $|\rho_3|$ are all nonzero. Then, using phase invariance of the three-wave system, we can normalize $\rho_1$, $\rho_2$ and $\rho_3$ to be all real. Thus, in the later text, we assume $(\rho_1, \rho_2, \rho_3)$ real. In addition, we define three real constants
\[ \label{gamma123}
\gamma_{1} \equiv \epsilon_1 \frac{ \rho_{2} \rho_{3} }{ \rho_{1}} ,\  \gamma_{2} \equiv \epsilon_2 \frac{\rho_{1} \rho_{3} }{ \rho_{2}}, \
\gamma_3\equiv \epsilon_3 \frac{ \rho_{1} \rho_{2} }{ \rho_{3}}.
\]

Rogue waves in the three-wave system (\ref{3WRIModel}) are solutions that arise from the above constant background, or from  non-constant solutions such as lumps or dark solitons that sit on the above constant background. Such solutions will be derived in later sections. Our solutions will be presented through elementary Schur polynomials $S_j(\mbox{\boldmath $x$})$, which are defined by
\begin{equation*}
\sum_{j=0}^{\infty}S_j(\mbox{\boldmath $x$})\epsilon^j
=\exp\left(\sum_{j=1}^{\infty}x_j\epsilon^j\right),
\end{equation*}
where $\mbox{\boldmath $x$}=(x_1,x_2,\cdots)$.

\section{Rogue waves that arise from a constant background}
Rogue waves that arise from the constant background (\ref{PlanewaveSolu2}) are rational solutions. Our general rational solutions to the three-wave system (\ref{3WRIModel}) are given in two different but equivalent forms by the following two theorems.

\begin{quote}
\textbf{Theorem 1.} The (2+1)-dimensional three-wave system (\ref{3WRIModel}) admits rational solutions
\begin{eqnarray}
  && q_{1,N}(x,y,t)= \rho_{1}\frac{g_{1,N}}{f_{N}} e^{i (k_{1}x + \lambda_{1}y + \omega_{1} t)}, \nonumber  \\
  && q_{2,N}(x,y,t)= \rho_{2}\frac{g_{2,N}}{f_{N}} e^{i (k_{2}x + \lambda_{2}y + \omega_{2} t)}, \label{diffopesolu2} \\
  && q_{3,N}(x,y,t)= \textrm{i} \rho_{3}\frac{g_{3,N}}{f_{N}} e^{\textrm{i} (k_{3}x + \lambda_{3} y + \omega_{3} t)}, \nonumber
\end{eqnarray}
where $N$ is an arbitrary positive integer,
\[ \label{diffopesolufN}
f_{N}=\tau_{0,0}, \quad g_{1,N}=\tau_{1,0}, \quad  g_{2,N}=\tau_{0,-1}, \quad g_{3,N}=\tau_{-1,1},
\]
\[ \label{deftaunk}
\tau_{n,k}=
\det_{
\begin{subarray}{l}
1\leq i, j \leq N
\end{subarray}
}
\left(
\begin{array}{c}
m_{i,j}^{(n,k)}
\end{array}
\right),
\]
the matrix elements in $\tau_{n,k}$ are defined by
\begin{eqnarray} \label{mij-diff}
&& m_{i,j}^{(n,k)}=
  \frac{\left(p\partial_{p}\right)^{n_{i}}}{ (n_{i}) !}\frac{\left(q \partial_{q}\right)^{n_{j}}}{(n_{j}) !} \left.
  \left[ \frac{1}{p + q}\left(-\frac{p}{q}\right)^{k}\left(-\frac{p-\textrm{i}}{q+\textrm{i}}\right)^{n} e^{\Theta_{i,j}(x,y,t)}\right]\ \right|_{p=p_{i}, \ q=p_{j}^*},
\end{eqnarray}
$n_{i}$ and $n_{j}$ are arbitrary non-negative integers,
\begin{eqnarray} \label{Thetaxt1}
\Theta_{i,j}(x,y,t)=  \left( \frac{1}{p} + \frac{1}{q} \right)z_1+  \left( \frac{1}{p-\textrm{i}} + \frac{1}{q+\textrm{i}} \right)z_2 + (p+q) z_3  + \sum _{r=1}^\infty  a_{r,i} \ln^r \left[ \frac{p}{p_{i}} \right] + \sum _{r=1}^\infty a^*_{r,j}  \ln^r \left[ \frac{q}{p_{j}^*} \right],
\end{eqnarray}
variables $(z_1, z_2, z_3)$ are related to $(x, y, t)$ as
\begin{eqnarray}
&& z_{1}=\gamma_1 \frac{V_{22}  x - V_{21} y}{V_{11} V_{22}-V_{12} V_{21}},   \label{defz1} \\
&& z_{2}=\gamma_2 \frac{V_{11} y-V_{12}  x}{V_{11} V_{22}-V_{12} V_{11}},     \label{defz2} \\
&& z_{3}=\gamma_3 \left[\frac{\left(V_{12}-V_{22}\right) x+\left(V_{21}-V_{11}\right)y}{V_{11} V_{22}-V_{12} V_{21}}+t\right],  \label{defz3}
\end{eqnarray}
$p_i$ are free non-imaginary complex constants, and $a_{r,i} \hspace{0.05cm} (r=1, 2, \dots)$ are free complex constants.
\end{quote}

One may notice from Eq.~(\ref{mij-diff}) that these solutions involve exponential factors $e^{\Theta_{i,j}}$. But those exponential factors will cancel out when $g_{i,N}$ is divided by $f_N$ in Eq.~(\ref{diffopesolu2}), and the resulting ratios are thus only rational functions of $(x, y, t)$.

The solutions in the above theorem are given through differential operators [see (\ref{mij-diff})]. More explicit expressions of these solutions are presented in the following theorem.
\begin{quote}
\textbf{Theorem 2.} The rational solutions in Theorem 1 are given by Eqs. (\ref{diffopesolu2})-(\ref{deftaunk}), where the matrix elements $m_{i,j}^{(n,k)}$ of $\tau_{n,k}$ can be rewritten as
\[ \label{Schmatrimnij}
m_{i,j}^{(n,k)}=\sum_{\nu=0}^{\min(n_{i}, n_{j})}\left(\frac{1}{p_{i}+p_{j}^*}\right) \left[ \frac{ p_{i} p_{j}^* }{(p_{i}+p_{j}^*)^2}  \right]^{\nu} \hspace{0.06cm} S_{n_{i}-\nu}[\textbf{\emph{x}}^{+}_{i,j}(n,k) +\nu \textbf{\emph{s}}_{i,j}] \hspace{0.06cm} S_{n_{j}-\nu}[\textbf{\emph{x}}^{-}_{j,i}(n,k) + \nu \textbf{\emph{s}}^*_{j,i}],
\]
vectors $\textbf{\emph{x}}^{\pm}_{i,j}(n,k)=\left( x_{1,i,j}^{\pm}, x_{2,i,j}^{\pm},\cdots \right)$ are
\begin{eqnarray}
&&x_{r,i,j}^{+}(n,k)= \alpha_{r, i}  z_1 +  \beta_{r, i} z_2 + \frac{1}{r!} p_{i} z_3 + n g_{r, i} + k h_{r} - c_{r, i,j} + a_{r, i},    \label{xrijplus}\\
&&x_{r,i,j}^{-}(n,k)= \alpha_{r, i}^* z_1 +  \beta_{r, i}^* z_2 + \frac{1}{r!} p_i^* z_3 - n g_{r,i}^* - k h_{r}^* - c_{r, i,j}^* + a_{r, i}^*,
\end{eqnarray}
$\textbf{\emph{s}}_{i,j}=(s_{1,i,j}, s_{2,i,j}, \cdots)$, coefficients
$\alpha_{r,i}$, $\beta_{r,i}$, $g_{r,i}$, $h_{r}$, $c_{r,i,j}$ and $s_{r,i,j}$ are obtained from the expansions
\begin{eqnarray}
&& \frac{1}{p_{i} e^{\kappa}}- \frac{1}{p_{i}} = \sum_{r=1}^{\infty} \alpha_{r,i}\kappa^{r}, \quad
\frac{1}{p_i e^{\kappa} -\rm{i}}-\frac{1 }{p_i-\rm{i}} = \sum_{r=1}^{\infty} \beta_{r,i}\kappa^{r},  \label{schucoefalpha}\\
&& \ln \left(\frac{p_i e^{\kappa} -\rm{i}}{p_i-\rm{i}} \right) =\sum_{r=1}^{\infty} g_{r,i}\kappa^{r}, \quad \kappa =\sum_{r=1}^{\infty} h_{r}\kappa^{r}, \\
&&  \ln \left[ \frac{p_i  e^{\kappa} +p_j^*}{p_i+p_j^*} \right]= \sum_{r=1}^{\infty}c_{r,i,j} \kappa^r, \ \ \ \ln \left[\frac{p_i + p_j^* }{\kappa} \left( \frac{ e^\kappa -1}{p_i e^\kappa + p_j^*} \right)  \right] = \sum_{r=1}^{\infty}s_{r,i,j} \kappa^r, \label{schurcoeffsr}
\end{eqnarray}
$p_i$ are free non-imaginary complex constants, and $a_{r,i} \hspace{0.05cm} (r=1, 2, \dots)$ are free complex constants.
\end{quote}

The above rational solutions contain rogue waves that arise from the constant background (\ref{PlanewaveSolu2}), as well as algebraic localized (lump) solitons moving on this constant background and a mixture between these two types of solutions (see later text). To get rogue waves that arise from the constant background, we need to impose conditions on $p_i$. To derive these conditions, we consider the fundamental rational solution, where we take $N=1$ and $n_{1}=1$ in Theorem 2. In addition, we normalize $a_{1,1}-c_{1,1,1}=0$ by a coordinate shift. Performing simple calculations, we can reduce the $\tau_{n,k}$ function to
\[  \label{fundamenRwsm11}
\tau_{n,k}=m_{1,1}^{(n,k)}= \xi \bar{\xi}+\Delta,
\]
where
\begin{eqnarray}
\xi=  a x + b y+ c t + \theta(n,k),\quad \bar{\xi}=  a^* x + b^* y+ c^* t - \theta^*(n,k), \quad \Delta= \frac{|p_1|^2}{(p_1+p_1^*)^2},
\end{eqnarray}
and $a$, $b$, $c$, $\theta$ are complex coefficients given by
\begin{eqnarray}
&& a=\frac{1}{V_{11} V_{22}-V_{12} V_{21}}\left[
-\frac{1}{p_1}\gamma_1 V_{22}+\frac{p_1}{(p_1-\textrm{i})^2}\gamma_2V_{12}+p_1\gamma_3(V_{12}-V_{22})\right], \label{complexcoeff1} \\
&& b=\frac{1}{V_{11} V_{22}-V_{12} V_{21}}\left[
\frac{1}{p_1}\gamma_1 V_{21}-\frac{p_1}{(p_1-\textrm{i})^2}\gamma_2V_{11}+p_1\gamma_3(V_{21}-V_{11})\right], \label{complexcoeff2} \\
&& c =\gamma_3 p_1,\ \ \  \theta(n,k) = k+\frac{n p_1}{p_1-\textrm{i}}. \label{complexcoeff3}
\end{eqnarray}

To derive parameter conditions for rogue waves, we separate the real and imaginary parts of the above complex variables as
\[ \label{complexcoeff4}
p_1=p_{1,1}+\textrm{i} p_{1,2},\  a=a_{1}+ \textrm{i} a_{2},\  b=b_{1}+ \textrm{i} b_{2}, \ c=c_{1}+ \textrm{i} c_{2},\ \theta=\theta_{1} + \textrm{i} \theta_{2}.
\]
Then, the $\tau_{n,k}$ solution (\ref{fundamenRwsm11}) can be rewritten as
\[
\tau_{n,k}= \xi_{1}^2+ \xi_{2}^2 - 2\textrm{i}\theta_{1}\xi_{2}+ 2\textrm{i}\theta_{2}\xi_{1}-\theta_{1}^2-\theta_{2}^2+\Delta,
\]
where
\[
\xi_{1}=a_{1} x + b_{1} y+ c_{1} t,\quad \xi_{2}=a_{2} x + b_{2} y+ c_{2} t.
\]
We can readily show that under the velocity assumption of $V_{11}V_{22}-V_{12}V_{21}\ne 0$, the three ratios of $a_1/a_2, b_1/b_2$ and $c_1/c_2$ cannot be all the same. Then, the above fundamental rational solution exhibits two distinctively different dynamical behaviours depending on the ratio relations between $a_1/a_2$ and $b_1/b_2$.
\begin{enumerate}
\item If $\frac{a_{1}}{a_{2}}\neq\frac{b_{1}}{b_{2}}$, then along the $\left[ x(t), y(t) \right]$ trajectory where
\[
a_{1} x + b_{1} y =-c_{1}t,\ \ \ a_{2} x + b_{2} y = -c_{2}t,
\]
the above $\tau_{n,k}$ is a constant. The corresponding solution $(q_1, q_2, q_3)$ is an algebraic localized lump soliton moving on the constant background (\ref{PlanewaveSolu2}).
\item If $\frac{a_{1}}{a_{2}}=\frac{b_{1}}{b_{2}}\ne \frac{c_1}{c_2}$, then the $\frac{a_{1}}{a_{2}}=\frac{b_{1}}{b_{2}}$ equation yields the following parameter condition
\begin{eqnarray} \label{condp11p12}
\frac{(p_{1,2}-1) (p_{1,1}^2+p_{1,2}^2)^2}{\gamma_1}-\frac{p_{1,2}[p_{1,1}^2+(p_{1,2}-1)^2]^2}{\gamma_2}
-\frac{p_{1,1}^2+p_{1,2}^2-p_{1,2}}{\gamma_3}=0.
\end{eqnarray}
In this case, the corresponding solution $(q_1, q_2, q_3)$ uniformly approaches the constant background (\ref{PlanewaveSolu2}) in the entire $(x,y)$ plane when $t\to\pm \infty$. In the intermediate times, it rises to a higher amplitude. Since $a_1/a_2=b_1/b_2$, this solution depends on $(x, y)$ through the combination of $a_{1} x + b_{1} y$. Thus, this is a line rogue wave.

For the other (i.e., non-fundamental) rational solutions in Theorems 1-2, in order for them to be rogue waves that arise from the constant background (\ref{PlanewaveSolu2}), we need to require every parameter $p_i$ $(1\le i\le N)$ to satisfy the above condition (\ref{condp11p12}), where $p_{1,1}$ is replaced by $p_{i,1}$ and $p_{1,2}$ replaced by $p_{i,2}$, with $(p_{i,1}, p_{i,2})$ being the real and imaginary parts of the complex parameter $p_i$.

The condition (\ref{condp11p12}) can be further simplified. When $(p_{1,1}, p_{1,2})$ are replaced by the more general $(p_{i,1}, p_{i,2})$, the simplified condition can be expressed as a quartic equation for $p_{i,1}$ as
\[\label{QuarticQinticEq}
\chi_{0}p_{i,1}^4 + \chi_{1} p_{i,1}^2 +\chi_{2}=0,
\]
where the coefficients are
\begin{eqnarray*}
&& \chi_{0}=(p_{i,2}-1)\gamma_1^{-1}-p_{i,2}\gamma_2^{-1}, \\
&& \chi_{1}=2 \left(p_{i,2}-1\right) p_{i,2}^2\gamma_1^{-1}-
2p_{i,2}\left(p_{i,2}-1\right)^2\gamma_2^{-1}
-\gamma_3^{-1}, \\
&& \chi_{2}=p_{i,2}\left(p_{i,2}-1\right) \left[p_{i,2}^3\gamma_1^{-1}-\left(p_{i,2}-1\right)^3\gamma_2^{-1}
-\gamma_3^{-1}\right].
\end{eqnarray*}
\end{enumerate}

\section{Dynamics of rogue waves that arise from a constant background}
Now we study dynamics of rogue waves that arise from the constant background (\ref{PlanewaveSolu2}) in the three-wave system (\ref{3WRIModel}). These solutions are given by Theorems 1-2, with all parameters $p_i$ satisfying the condition (\ref{QuarticQinticEq}).

\subsection{Fundamental rogue waves}
The fundamental rogue wave is given by Eq.~(\ref{fundamenRwsm11}), with $p_1$ satisfying the condition (\ref{QuarticQinticEq}). The amplitude functions of this fundamental rogue wave can be written more explicitly as
\[ \label{Fundamplitu1}
|q_{i,1}(x,t)|=\left| \rho_{i} \frac{g_{i,1}}{f_{1}} \right|, \quad i=1,2,3,
\]
where
\begin{eqnarray*}
&& f_{1}(x,y,t) =m_{1,1}^{(0,0)}=(a_{1} x + b_{1} y+ c_{1} t)^2+  \frac{1}{\zeta_{0}^2}(a_{1} x + b_{1} y+ \zeta_{0} c_{2} t)^2+\frac{|p_1|^2}{(p_1+p_1^*)^2},\\
&& g_{1,1}(x,y,t)= m_{1,1}^{(1,0)}= f_{1}(x,y,t)-  \frac{2i \hat{\theta}_{1}}{\zeta_{0}}\left( a_{1} x + b_{1} y+ \zeta_{0} c_{2} t \right)+2i \hat{\theta}_{2}(a_{1} x + b_{1} y+ c_{1} t)-\hat{\theta}_{1}^2-\hat{\theta}_{2}^2, \\
&& g_{2,1}(x,y,t)= m_{1,1}^{(0,-1)}= f_{1}(x,y,t)  + \frac{2i}{\zeta_{0}}\left(a_{1} x + b_{1} y+ \zeta_{0} c_{2}   t\right)-1, \\
&& g_{3,1}(x,y,t)= m_{1,1}^{(-1,1)}=f_{1}(x,y,t)+\frac{2i (\hat{\theta}_{1}-1)}{\zeta_{0} }  \left(a_{1} x + b_{1} y+  \zeta_{0} c_{2} t\right)
-2i \hat{\theta}_{2}(a_{1} x + b_{1} y+ c_{1} t)-(\hat{\theta}_{1}-1)^2-\hat{\theta}_{2}^2,
\end{eqnarray*}
parameters $(a_1, b_1, c_1, c_2)$ are given by Eqs. (\ref{complexcoeff1})-(\ref{complexcoeff4}), and
\[
\zeta_{0}=\frac{a_{1}}{a_{2}}, \quad \hat{\theta}_{1}=\Re\left( \frac{p}{p-\rm{i}}\right), \quad  \hat{\theta}_{2}=\Im\left(\frac{p}{p-\rm{i}}\right).
\]
Here, $\Re$ and $\Im$ represent the real and imaginary parts of a complex number.

\begin{figure}[htb]
\begin{center}
\vspace{4.00cm}
\includegraphics[scale=0.550, bb=200 0 385 620]{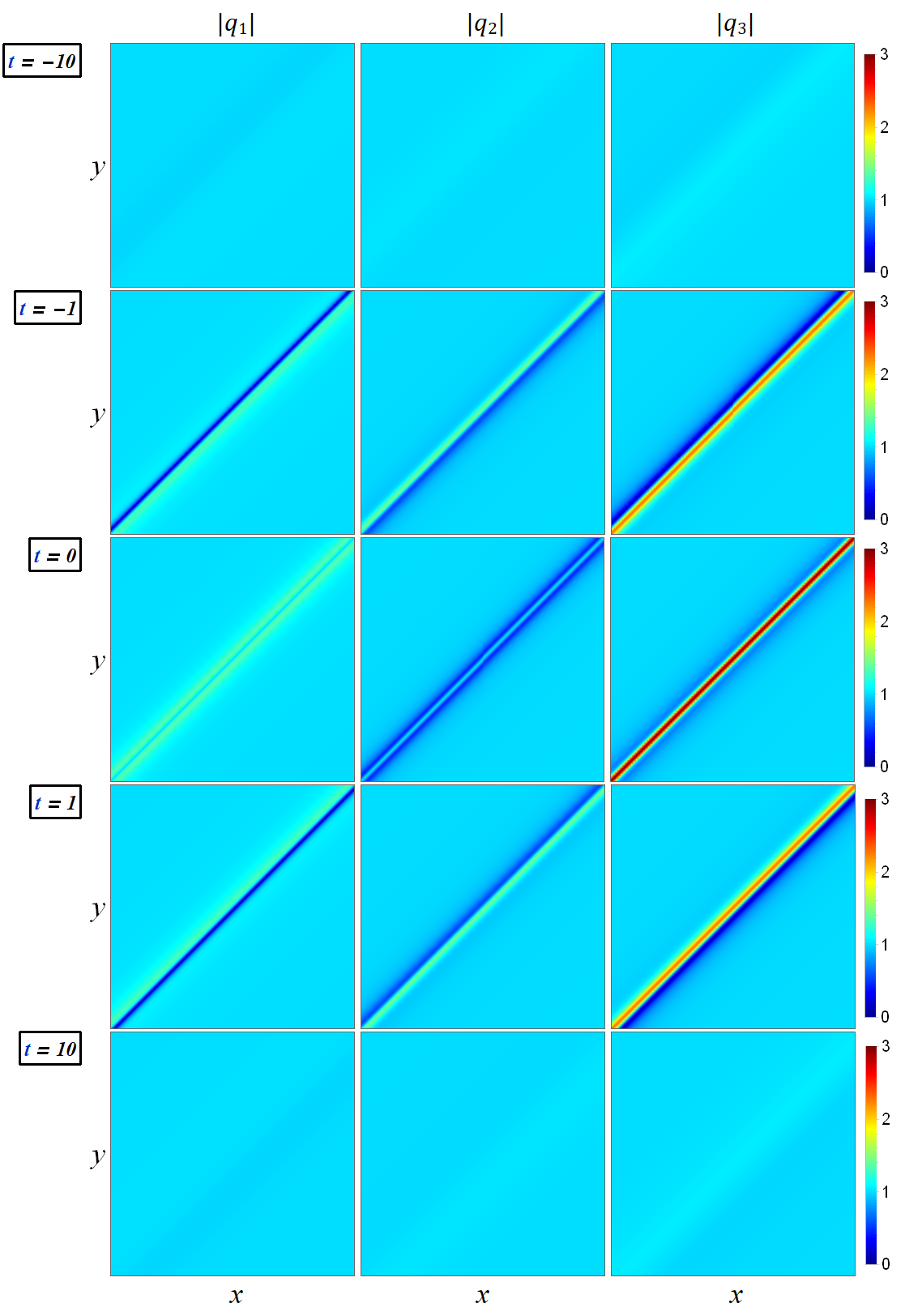}
\caption{A fundamental line rogue wave with parameter choices (\ref{DoubleRootpara1}) and $p=0.5+0.5\rm{i}$. In all panels, $ -10\le x, y\le 10$. }
\end{center}
\end{figure}

To demonstrate this fundamental rogue wave, we choose nonlinear coefficients, background amplitudes and velocity values as
\[\label{DoubleRootpara1}
\epsilon_1=-\epsilon_2=\epsilon_3=1, \
\rho_{1}=\rho_{2}=\rho_3=1,\ V_{1,1}=6, \ V_{1,2}=5,\ V_{2,1}=4,\ V_{2,2}=3.
\]
In addition, we choose $p_1=0.5+ 0.5 \rm{i}$, which satisfies the condition (\ref{QuarticQinticEq}).
For these choices of parameters, the corresponding rogue wave is displayed in Fig.~1. It is seen that this is a line rogue wave with a single dominant peak line.

\subsection{Multi-rogue waves}
To get multi-rogue waves, we set $N>1$ and $n_{1}= n_2=\cdots =n_N =1$ in the rational solutions of Theorem 2, and require all $\{p_i\}$ values to satisfy condition (\ref{QuarticQinticEq}). To demonstrate, we choose $N=2$ and the same nonlinear coefficients, background amplitudes and velocity values as in (\ref{DoubleRootpara1}). In addition, we choose
\[ \label{p1p2fig2}
p_{1}= -0.539770966 +0.3 \textrm{i}, \quad p_{2}=1.904662796 + 0.65 \textrm{i}, \quad a_{1, 1}=0, \quad a_{1,2}=0.
\]
Notice that these $(p_1, p_2)$ values satisfy conditions (\ref{QuarticQinticEq}), because we obtained their real parts by solving the quartic equation (\ref{QuarticQinticEq}), with their imaginary parts set as $0.3$ and $0.65$ respectively.
The corresponding two-rogue wave solution is displayed in Fig.~2. We see that this rogue wave features two intersecting lines,
indicating that this solution describes the interaction between two fundamental line rogue waves. Interestingly, at the intersection point between the two line rogue waves, the $|q_2|$ component has higher amplitude, but the $(|q_1|, |q_3|)$ components have lower amplitudes.

\begin{figure}[htb]
\begin{center}
\vspace{4.00cm}
\includegraphics[scale=0.550, bb=200 0 385 620]{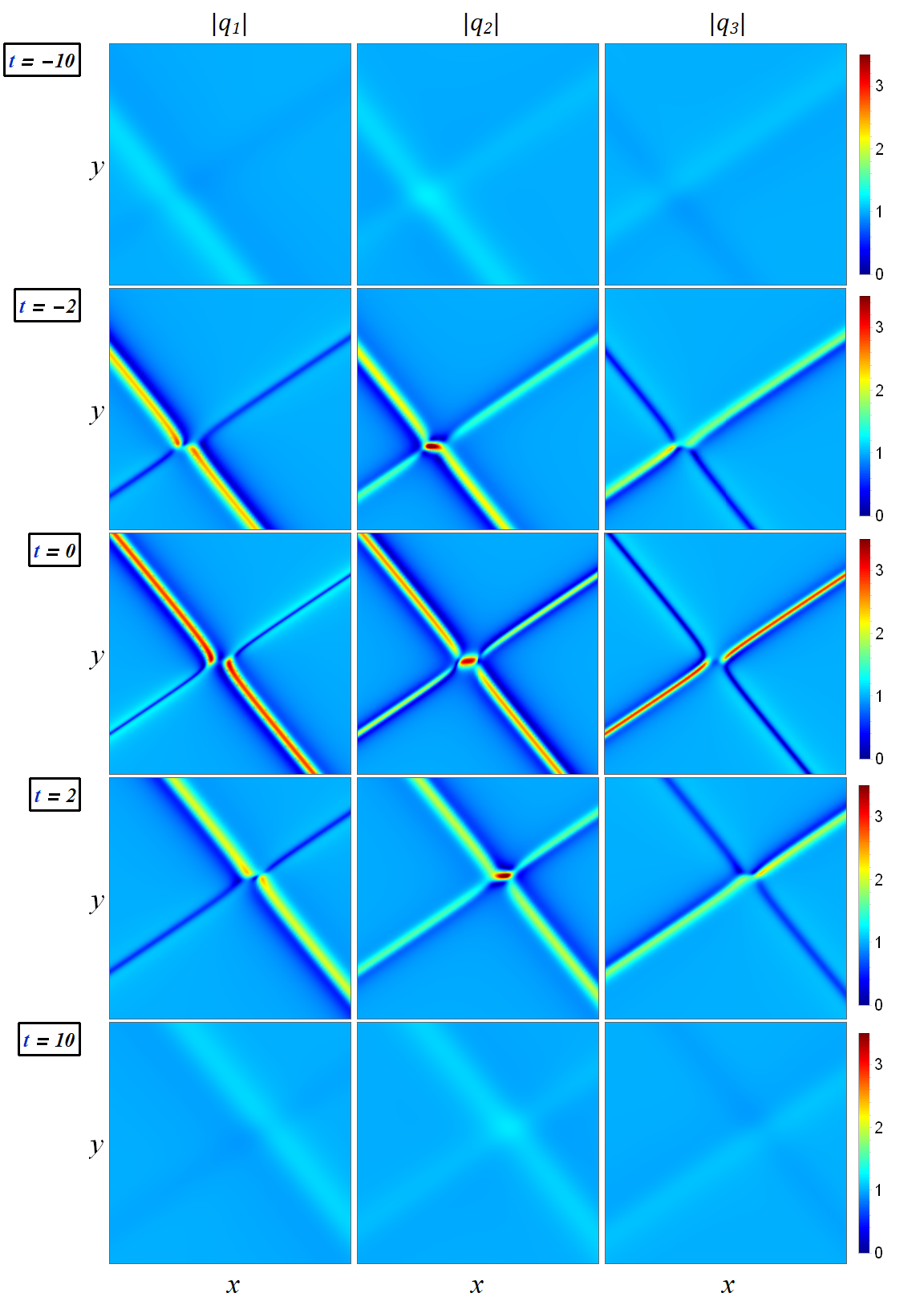}
\caption{A two-rogue wave solution with parameter choices (\ref{DoubleRootpara1})-(\ref{p1p2fig2}). In these panels, $-100\le x\le 0$ and $ -50\le y\le 0 $ in the first row, $-50\le x\le 50$ and $ -25\le y\le 25 $ in the second to fourth rows, and $0\le x\le 100$, $ 0\le y\le 50 $ in the last row. These $(x, y)$ intervals in different rows are different because the intersection point of the two line rogue waves is moving. }
\end{center}
\end{figure}

\subsection{High-order rogue waves}
Higher-order rogue waves are obtained by taking $N=1$ and $n_{1}>1$ in the rational solutions of Theorem 2, with $p_1$ satisfying the condition (\ref{QuarticQinticEq}). To demonstrate, we choose $n_1=2$, $p_1=\sqrt{3}/2+\textrm{i}/2$, and
\[\label{DoubleRootpara3}
\epsilon_1=-\epsilon_2=\epsilon_3=1, \
\rho_{1}=1,\ \rho_{2}=\sqrt{2},\ \rho_{3}=1,\ V_{1,1}=6, \ V_{1,2}=5,\ V_{2,1}=4,\ V_{2,2}=3, \ a_{1,1}=0, \ a_{2,1}=0.
\]
Note that for this $p_1$ value, $\sqrt{3}/2$ is a double root of the quartic equation (\ref{QuarticQinticEq}). The corresponding solution is displayed in Fig.~3. Interestingly, this second-order rogue solution is still a line rogue wave. However, compared to the single-peak fundamental line rogue wave of Fig.~1, this second-order line rogue wave develops two peaks, which are most pronounced in the $t=-1$ panels. This type of higher-order rogue waves has not been reported before to our best knowledge.

\begin{figure}[htb]
\begin{center}
\vspace{4.00cm}
\includegraphics[scale=0.550, bb=200 0 385 620]{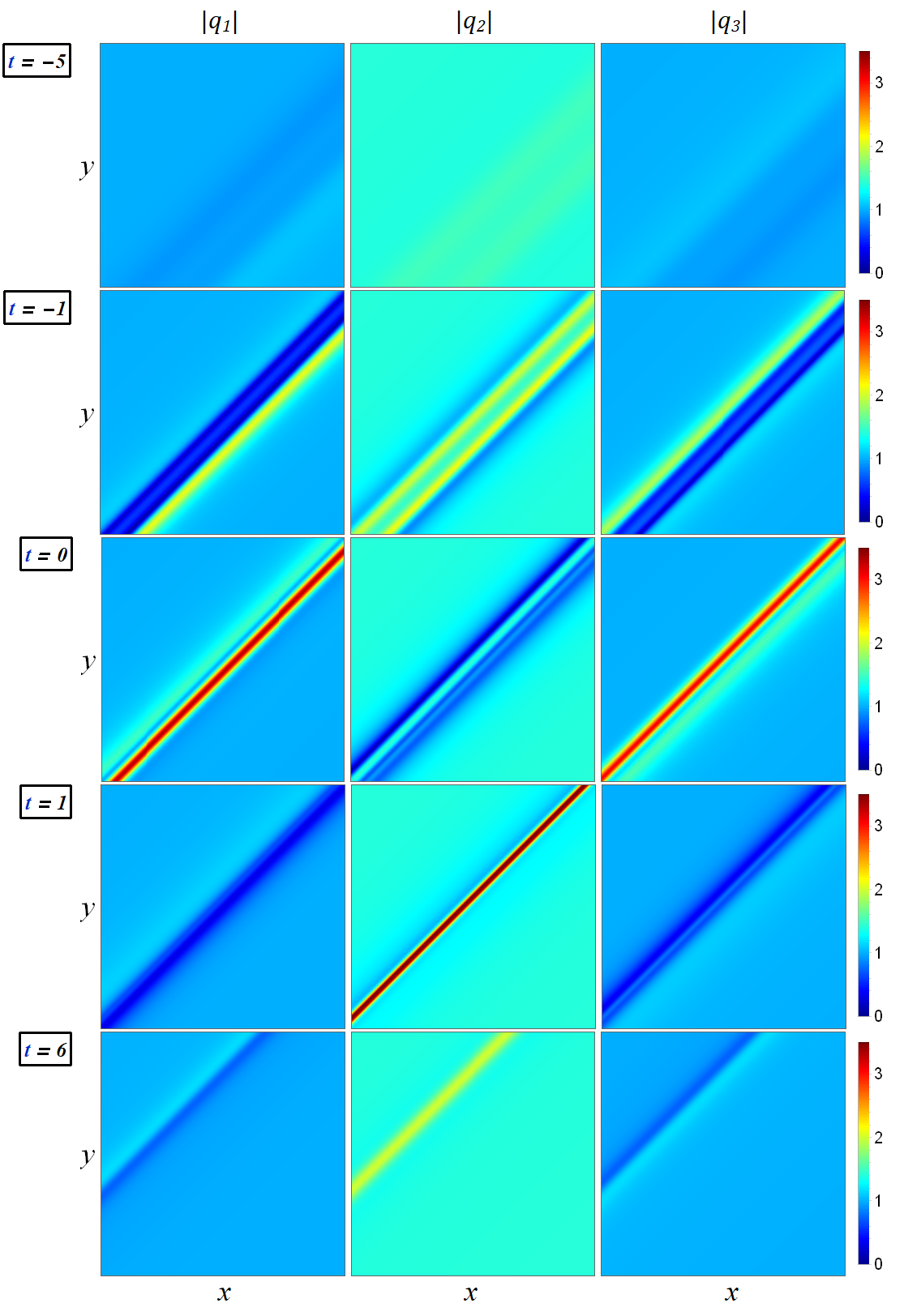}
\caption{A second-order rogue wave under parameter choices (\ref{DoubleRootpara3}) and
$p_{1}=\sqrt{3}/2+\textrm{i}/2$. In all panels, $-16\le x\le 16$ and $-8\le y \le 8$. }
\end{center}
\end{figure}

\section{Rogue waves that arise from a non-constant background}
The three-wave system (\ref{3WRIModel}) also admits rogue waves that do not arise from the
constant background (\ref{PlanewaveSolu2}). We consider this type of rogue waves in this section.

\subsection{Rogue waves that arise from lump-soliton backgrounds}
In the previous rational solutions of Theorems 1-2, if only some of the $\{p_i\}$ parameters satisfy the condition (\ref{QuarticQinticEq}) but the others do not, then we will get rogue waves that arise from lump-soliton backgrounds. To demonstrate this type of rogue waves, we choose
\[ \label{Parafig4}
N=2, \quad n_1=n_2=1, \quad p_1=2, \quad p_2=1+{\rm{i}}, \quad a_{1, 1}=0, \quad a_{1,2}=0,
\]
and the other parameters are as given in Eq.~(\ref{DoubleRootpara1}). Notice that $p_1$ does not satisfy the condition (\ref{QuarticQinticEq}), but $p_2$ does. The corresponding solution is displayed in Fig.~4. Notice that at large negative time, the solution is a lump soliton moving on the constant background (\ref{PlanewaveSolu2}). However, as time increases, a line rogue wave arises and interacts with this lump soliton. This interaction creates interesting intensity patterns which strongly deform the original lump soliton. At large positive time, however, this rogue wave disappears, and the solution goes back to a lump soliton again.

\begin{figure}[htb]
\begin{center}
\vspace{4.00cm}
\includegraphics[scale=0.550, bb=200 0 385 620]{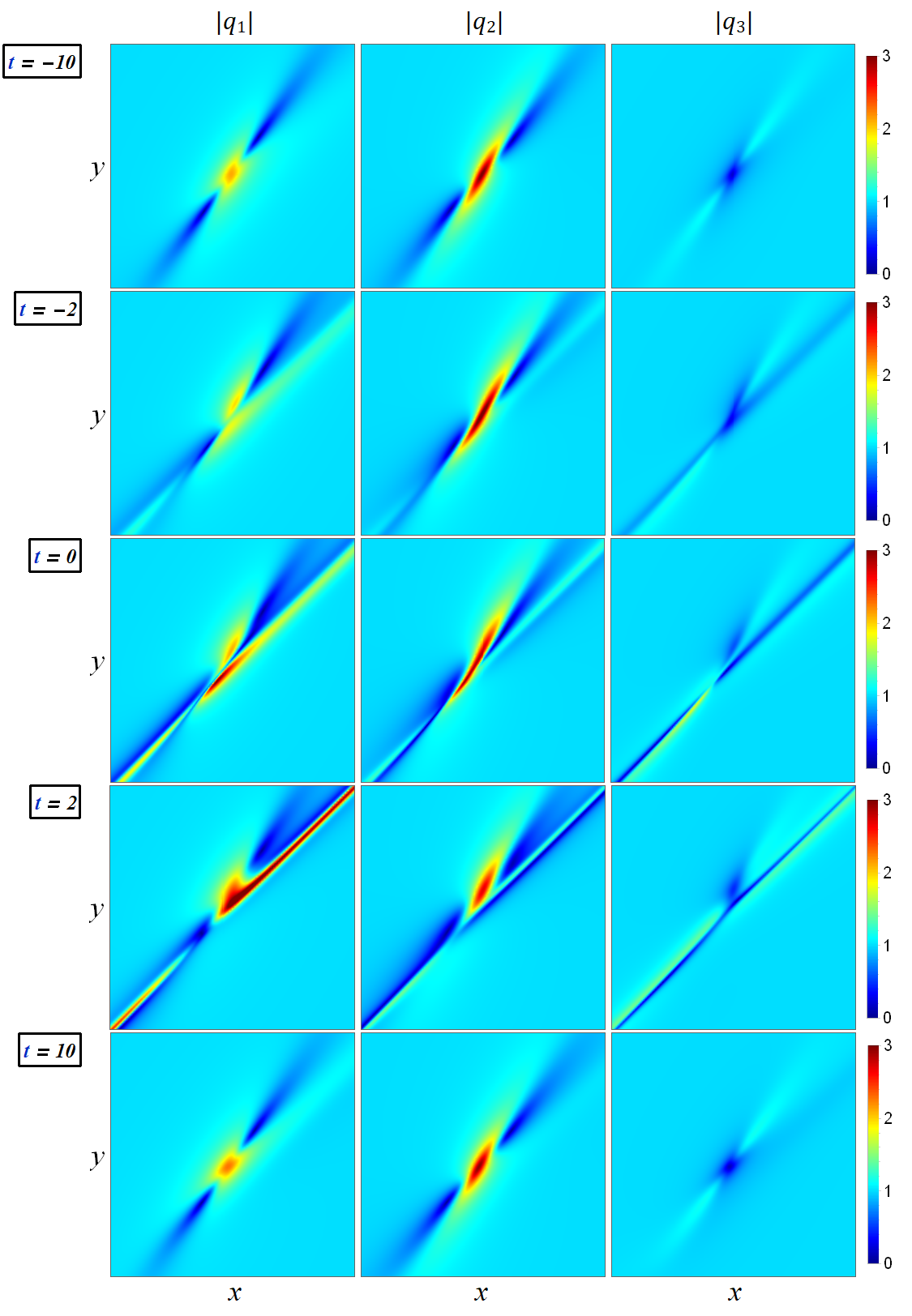}
\caption{A rogue wave that arises from a lump-soliton background. Solution parameters are given in Eqs.~(\ref{DoubleRootpara1}) and (\ref{Parafig4}). In these panels, $-56\le x\le -40 $ and $ -45\le y\le -33 $ in the first row, $-18\le x\le -2$ and $ -2\le y\le -14 $ in the second row, $-8\le x\le 8$ and $ -6\le y\le 6 $ in the third row, $2\le x\le 18$ and $ 2\le y\le 14$ in the fourth row, and $41\le x\le 57$, $ 36\le y\le 48 $ in the last row. These $(x, y)$ intervals in different rows are different because the lump soliton is moving. }
\end{center}
\end{figure}

\subsection{Rogue waves that arise from line-dark-soliton backgrounds}
In addition to the above rogue waves that arise from lump-soliton backgrounds, there also exist rogue waves that arise from line-dark-soliton backgrounds. This latter type of rogue waves are not rational solutions though. They are semi-rational solutions which contain both rational and exponential components. Analytical expressions of these semi-rational solutions are given by the following theorem.

\begin{quote}
\textbf{Theorem 3.} The (2+1)-dimensional three-wave system (\ref{3WRIModel}) admits semi-rational solutions (\ref{diffopesolu2})-(\ref{deftaunk}), where the matrix elements $m_{i,j}^{(n,k)}$ of $\tau_{n,k}$ are given by
\begin{eqnarray} \label{mij-diff5}
&& m_{i,j}^{(n,k)}=\pi_{i,j}+
  \frac{\left(p\partial_{p}\right)^{n_{i}}}{ (n_{i}) !}\frac{\left(q \partial_{q}\right)^{n_{j}}}{(n_{j}) !} \left.
  \left[ \frac{1}{p + q}\left(-\frac{p}{q}\right)^{k}\left(-\frac{p-\textrm{i}}{q+\textrm{i}}\right)^{n} e^{\Theta_{i,j}(x,y,t)}\right]\ \right|_{p=p_{i}, \ q=p_{j}^*},
\end{eqnarray}
or equivalently, by
\begin{eqnarray} \label{Schmatrimncij}
&& m_{i,j}^{(n,k)}=\left[\sum_{\nu=0}^{\min(n_{i}, n_{j})} \left(\frac{1}{p_{i}+p_j^*}\right) \left[ \frac{ p_{i} p_j^* }{(p_{i}+p_j^*)^2}  \right]^{\nu} \hspace{0.06cm} S_{n_{i}-\nu}[\textbf{\emph{x}}^{+}_{i,j}(n,k) +\nu \textbf{\emph{s}}_{i,j}] \hspace{0.06cm} S_{n_{j}-\nu}[\textbf{\emph{x}}^{-}_{j,i}(n,k) + \nu \textbf{\emph{s}}^*_{j,i}\right]   \nonumber  \\
&&    +\pi_{i,j} \left(-\frac{p_{i}}{p_j^*}\right)^{-k}\left(-\frac{p_{i}-\textrm{i}}{p_j^*+\textrm{i}}\right)^{-n} \exp\left[-\left( \frac{1}{p_{i}}+ \frac{1}{p_j^*} \right)z_1 - \left( \frac{1}{p_{i}-\textrm{i}} + \frac{1}{p_j^*+\textrm{i}} \right)z_2 - (p_{i}+p_j^*) z_3 \right],
\end{eqnarray}
$\pi_{i,j}$ are complex constants satisfying the constraint
\[  \label{condpiij}
\pi_{i,j}=\pi^*_{j,i},
\]
and all other notations are the same as in Theorems 1 and 2.
\end{quote}

If all $\pi_{i,j}$ are zero, then the above solution reduces to the rational solutions given in Theorems 1-2. If $n_i=0$ for all $1\le i\le N$, where the summation term in the above equation (\ref{Schmatrimncij}) reduces to a constant $1/(p_{i}+p_j^*)$, we would get line dark solitons or multi-line dark solitons. But if $\pi_{i,j}$ are not all zero, and $n_i$ are not all zero, then we will get semi-rational solutions.

Rogue waves will be obtained if some of the $p_i$ parameters in these semi-rational solutions satisfy the condition (\ref{QuarticQinticEq}), and these rogue waves will arise from single-line or multi-line dark solitons. We illustrate this type of rogue waves below.

Fundamental semi-rational rogue waves are obtained when we take $N=1$, $n_1=1$, $p_1$ satisfying the condition (\ref{QuarticQinticEq}), and $\pi_{1,1}$ is a real nonzero constant [hence satisfying the constraint (\ref{condpiij})]. To illustrate this type of rogue waves, we choose
\[ \label{Parafig5}
p_1=0.536234874+0.4 {\rm{i}}, \quad \pi_{1,1}=1, \quad a_{1,1}=0,
\]
and all other parameters are as given in Eq.~(\ref{DoubleRootpara1}). The corresponding solution is displayed in Fig.~5. As is seen, at large negative time, this solution is a fundamental line dark soliton. As time increases, a rogue wave appears, which interacts with this dark soliton. The most interesting feature of this rogue wave is that it is a half-line rogue wave which appears on only the upper side of the line dark soliton. This contrasts the fundamental rogue wave arising from a constant background in Fig.~1, which is a whole line. At large positive time, this half-line rogue wave disappears, and the solution goes back to the fundamental line dark soliton.

\begin{figure}[htb]
\begin{center}
\vspace{4cm}
\includegraphics[scale=0.550, bb=200 0 385 620]{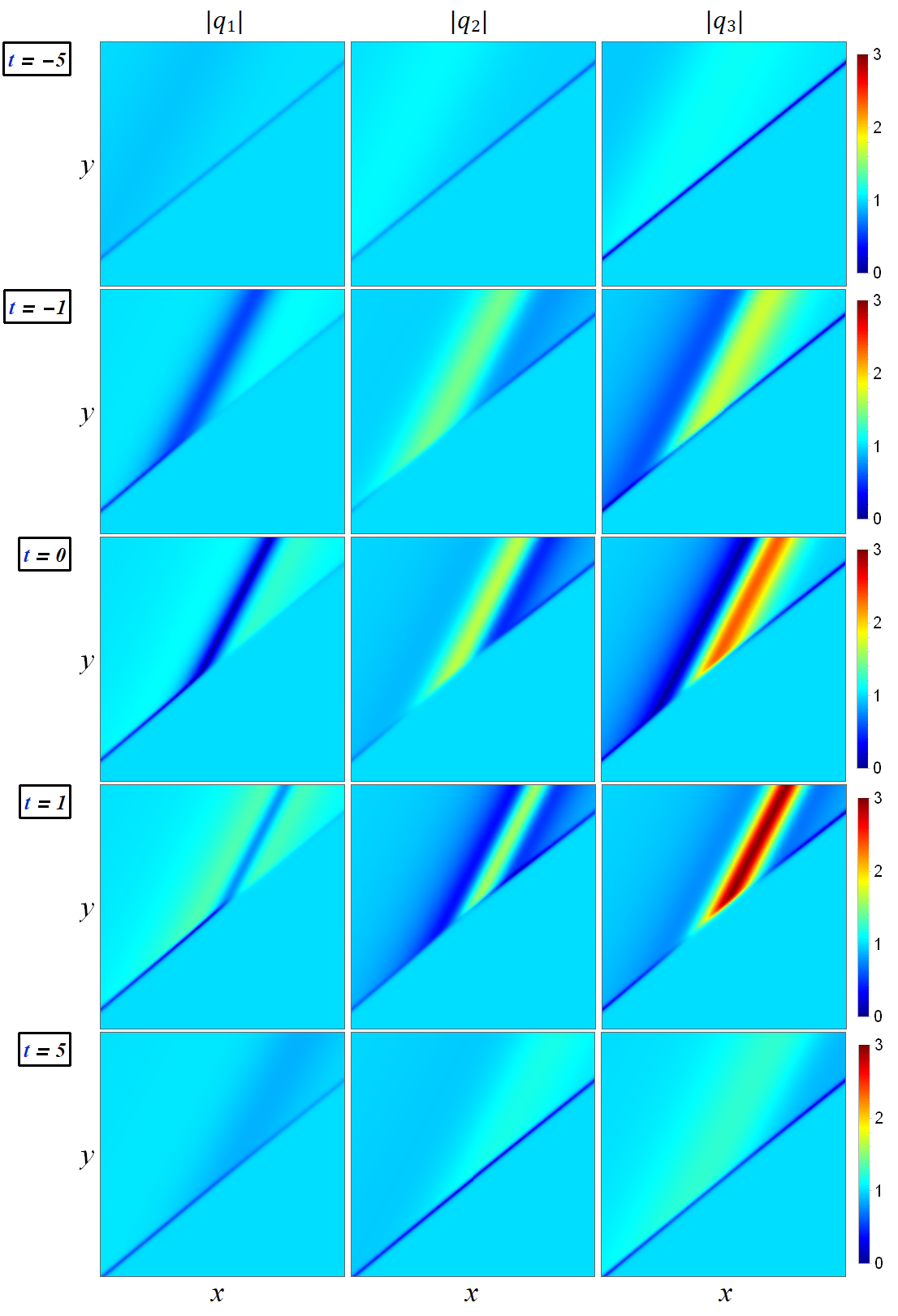}
\caption{A fundamental semi-rational rogue wave with parameter values in Eqs.~(\ref{DoubleRootpara1}) and (\ref{Parafig5}).
In these panels, $-8\le x, y\le 8 $ in the first four rows, and $-3\le x, y \le 13$ in the last row. }
\end{center}
\end{figure}

To demonstrate non-fundamental semi-rational rogue waves, we take
\[ \label{Parafig67}
N=2, \quad n_1=n_2=1, \quad \pi_{2,2}=1, \quad \pi_{1,2}=\pi_{2,1}=0, \quad a_{1,1}=0, \quad a_{1,2}=0.
\]
The nonlinear coefficients, background amplitudes and velocities are taken as in Eq.~(\ref{DoubleRootpara1}). We will show two solutions.

In the first non-fundamental semi-rational rogue wave, we choose
\[ \label{Parafig6}
p_1=-0.542592881+0.35{\rm i}, \quad p_2=1.904662796 + 0.65 \rm{i}, \quad \pi_{1,1}=-1.
\]
Both $p_1$ and $p_2$ here satisfy the condition (\ref{QuarticQinticEq}). Thus, we anticipate that this solution describes the interaction between a two-rogue-wave solution from a constant background and a two-dark-soliton solution. This solution is plotted in Fig.~6. It is seen that at large negative and positive times, this solution is indeed a two-dark soliton, and at intermediate times, a rogue wave appears and interacts with this two-dark soliton. These behaviors are consistent with our anticipations. However, this rogue wave that appears at intermediate times here is not a two-rogue wave from a constant background as the one shown in Fig.~2 (which features two intersecting whole lines). Instead, the present rogue wave consists of two half-lines which appear on the opposite sides of the two-dark soliton. In addition, locations of these two half-lines are also moving along the dark-soliton lines. The half-line feature of the present rogue wave echoes that in Fig.~5, but motions of the two half-line rogue waves (especially relative to each other) are new and fascinating features.

\begin{figure}[htb]
\begin{center}
\vspace{4.00cm}
\includegraphics[scale=0.550, bb=200 0 385 590]{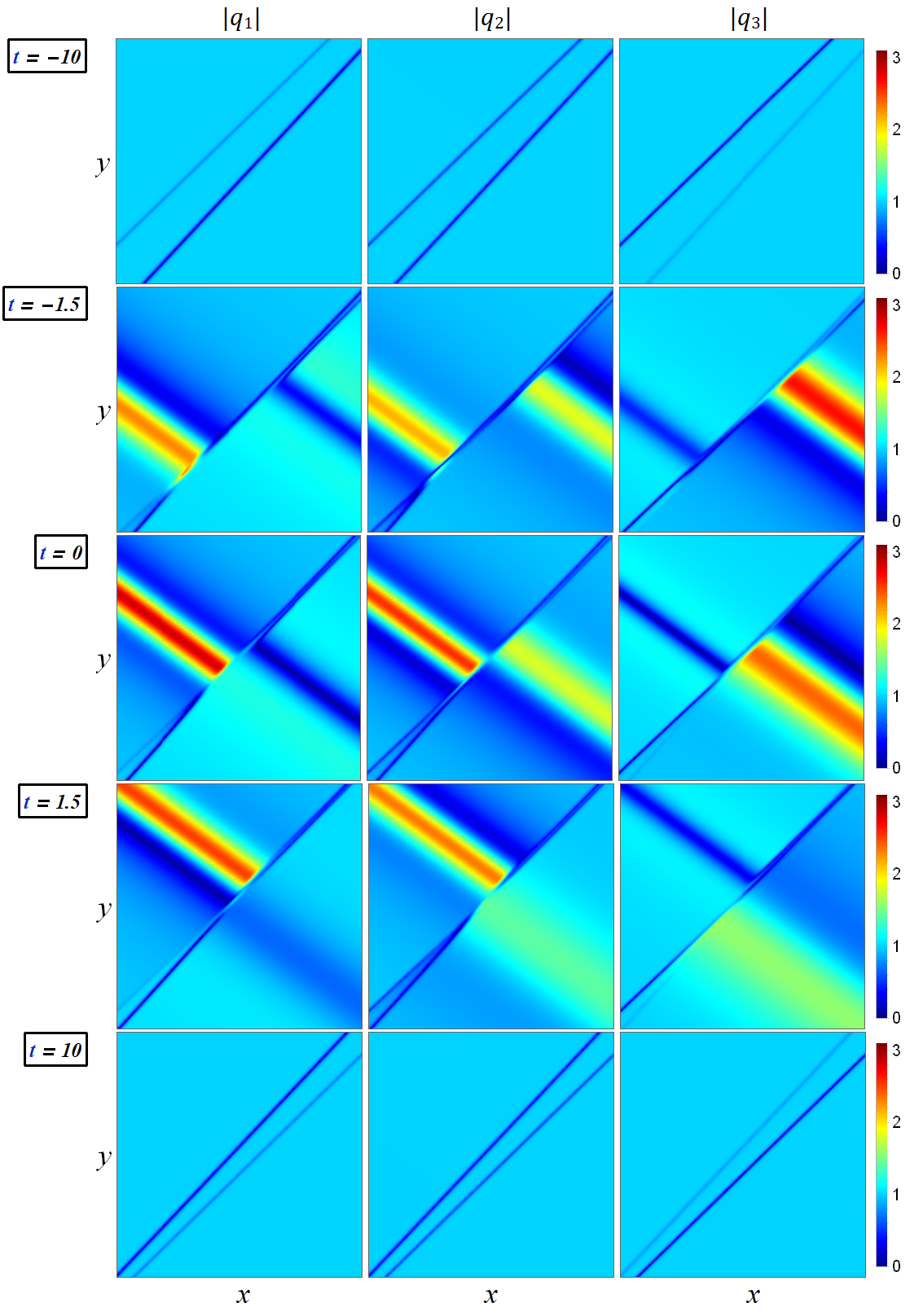}
\caption{A non-fundamental semi-rational rogue wave with parameter values given in Eqs.~(\ref{DoubleRootpara1}) and (\ref{Parafig67})-(\ref{Parafig6}). In these panels, $-97\le x\le -73$ and $-80\le y\le -60$ in the first row, $-17\le x\le 7$ and $ -13\le y\le 7 $ in the second row, $-12\le x\le 12$ and $ -10\le y\le 10 $ in the third row, $-7\le x\le 17$ and $ -7\le y\le 13$ in the fourth row, and $ 73\le x\le 97$, $ 60\le y\le 80 $ in the last row. }
\end{center}
\end{figure}

In the second non-fundamental semi-rational rogue wave, we choose
\[ \label{Parafig7}
p_1=1.5, \quad p_2=1+{\rm i}, \quad \pi_{1,1}=1.
\]
The $p_1$ value here does not satisfy the condition (\ref{QuarticQinticEq}), but $p_2$ does. Intuitively, this $p_1$ value in the rational part of the semi-rational solution signals a lump soliton, and the $p_2$ value signals a rogue wave --- likely a half-line rogue wave in view of Figs.~6 and 7. Thus, we might have anticipated that this semi-rational solution would describe the interaction between a lump soliton, a half-line rogue wave and a two-dark soliton. The true solution is displayed in Fig.~7. One of the most surprising features of this solution is that, at large negative time, the lump soliton is absent. However, as time increases, this lump miraculously appears. More interestingly, as time increases further, this lump persists but detaches itself and moves away from the two-dark soliton. In the whole evolution process, there is no half-line rogue wave. Instead, the lump that emerges from the two-dark soliton acts as a rogue wave. But this lump rogue wave does not disappear at large positive time, unlike rogue waves in previous six figures.

\begin{figure}[htb]
\begin{center}
\vspace{4.00cm}
\includegraphics[scale=0.550, bb=200 0 385 590]{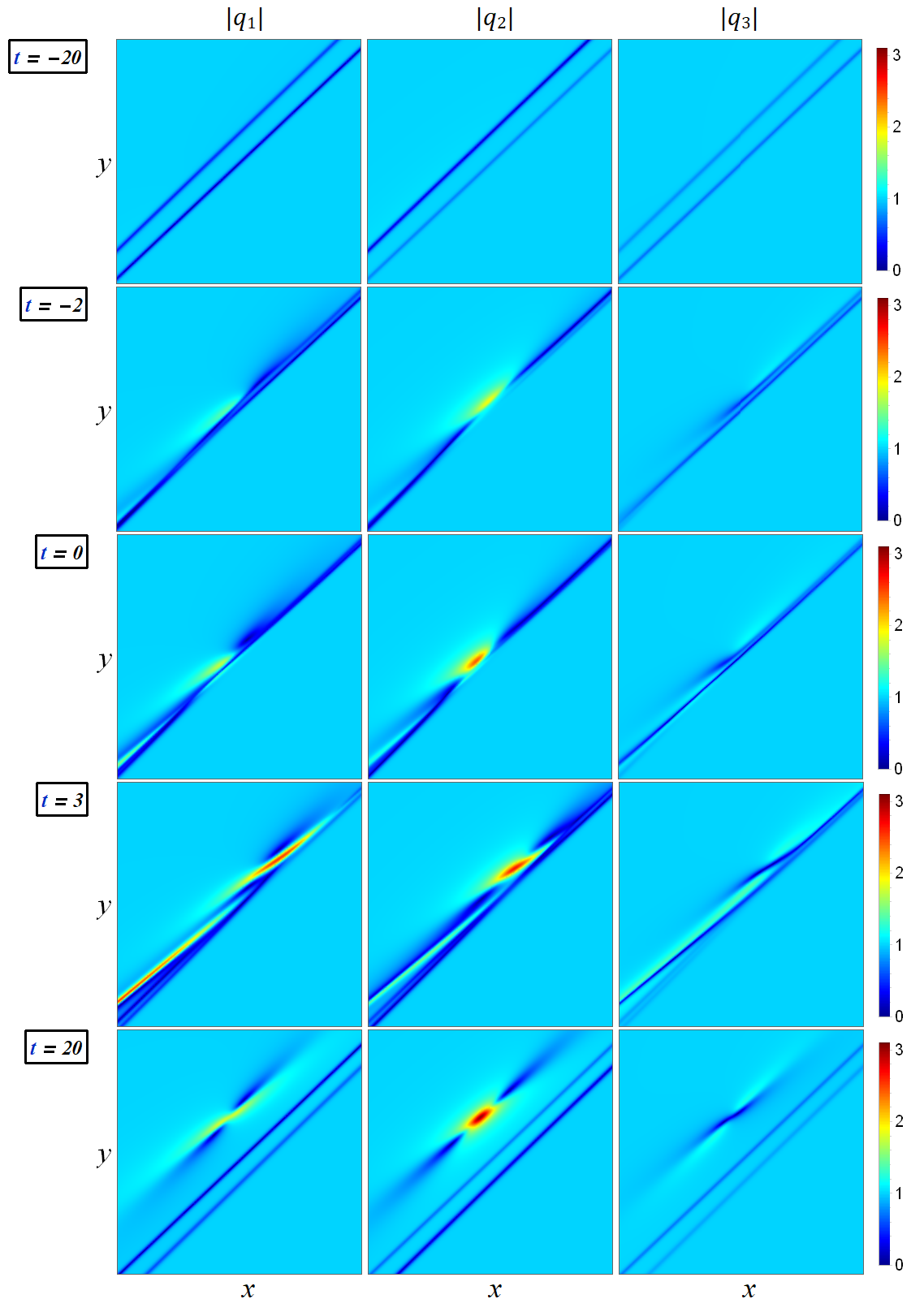}
\caption{Another non-fundamental semi-rational rogue wave with parameter values given in Eqs.~(\ref{DoubleRootpara1}), (\ref{Parafig67}) and (\ref{Parafig7}). In these panels, $-92\le x\le -72 $ and $ -75\le y\le -57 $ in the first row, $-20\le x\le 0$ and $ -17\le y\le 1 $ in the second row, $-10\le x\le 10$ and $ -9\le y\le 9 $ in the third row, $0\le x\le 20$ and $ -1\le y\le 17$ in the fourth row, and $ 72\le x\le 92$, $ 57\le y\le 75 $ in the last row. }
\end{center}
\end{figure}

\section{Proofs of Theorems 1-3}\label{sec:derivation}
In this section, we derive the rational and semi-rational solutions presented in Theorems 1-3.

\subsection{Proof of Theorem 1}
Due to the boundary conditions (\ref{PlanewaveSolu2}), we first introduce the bilinear transformation
\begin{eqnarray}
  && q_{1}(x,y,t)= \rho_{1}\frac{g_{1}}{f} e^{i (k_{1}x + \lambda_{1}y + \omega_{1} t)}, \nonumber \\
  && q_{2}(x,y,t)= \rho_{2}\frac{g_{2}}{f} e^{i (k_{2}x + \lambda_{2}y + \omega_{2} t)}, \label{Transform} \\
  && q_{3}(x,y,t)= \textrm{i} \rho_{3}\frac{g_{3}}{f} e^{\textrm{i} (k_{3}x + \lambda_{3} y + \omega_{3} t)}, \nonumber
\end{eqnarray}
where $f(x,y,t)$ is a real function, and $g_k(x,y,t) \hspace{0.05cm} (k=1, 2, 3)$ are complex functions. Under this transformation, the three-wave system (\ref{3WRIModel}) is converted into the following system of bilinear equations
\begin{eqnarray}
\left( D_{t}+ V_{1,1} D_{x}+ V_{1,2} D_{y}  -{\rm{i}}  \gamma_1 \right) g_{1}  \cdot f & = & -{\rm{i}}  \gamma_1 g_{2}^* g_{3}^*, \nonumber  \\
\left( D_{t}+V_{2,1} D_{x} + V_{2,2} D_{y}-{\rm{i}}  \gamma_2 \right) g_{2}  \cdot f & = & -{\rm{i}}  \gamma_2 g_{1}^* g_{3}^*, \label{1stBilineform2}  \\
\left( D_{t} -{\rm{i}}  \gamma_3  \right) g_{3}  \cdot f & = & -{\rm{i}}  \gamma_3  g_{1}^* g_{2}^*,  \nonumber
\end{eqnarray}
where $D$ is Hirota's bilinear differential operator defined by
\begin{eqnarray*}
P \left(D_{x}, D_{y}, D_{t}\right) F(x,y,t) \cdot G(x,y,t) \equiv P\left(\partial_{x}-\partial_{x'}, \partial_{y}-\partial_{y'}, \partial_{t}-\partial_{t'}\right) F(x,y,t) G(x',y',t')|_{x'=x, \hspace{0.05cm} y'=y, \hspace{0.05cm} t'=t},
\end{eqnarray*}
$P$ is a polynomial of $(D_{x}, D_{y}, D_{t})$, and constants $(\gamma_1, \gamma_2, \gamma_3)$ are as defined in Eq.~(\ref{gamma123}).

Next, we introduce the coordinate transformation
\[ \label{VariaTrans1}
\left( \begin{array}{ccc}
x \\
y \\
t \\
\end{array} \right)  =
\left(
\begin{array}{ccc}
\frac{V_{1,1}}{\gamma_{1}} & \frac{V_{2,1}}{\gamma_{2}} & 0  \\
\frac{V_{1,2}}{\gamma_{1}}  & \frac{V_{2,2}}{\gamma_{2}} & 0 \\
\frac{1}{\gamma_{1}}  &  \frac{1}{\gamma_{2}} & \frac{1}{\gamma_{3}}  \\
\end{array}
\right)\left( \begin{array}{ccc}
z_{1} \\
z_{2}   \\
z_{3}\\
\end{array} \right),
\]
which is equivalent to Eqs. (\ref{defz1})-(\ref{defz3}) in Theorem 1.
Under this coordinate transformation, we have
\begin{eqnarray}
&&  \partial_{t} +  \mathbf{V}_{1} \cdot \nabla = \gamma_1 \partial_{z_{1}}, \nonumber  \\
&&  \partial_{t} + \mathbf{V}_{2} \cdot \nabla  = \gamma_2 \partial_{z_{2}},  \label{partialtrans2}  \\
&&  \partial_{t}                                = \gamma_3 \partial_{z_{3}}.   \nonumber
\end{eqnarray}
Thus, the bilinear system (\ref{1stBilineform2}) reduces to
\begin{eqnarray}
&& \left( {\rm{i}} D_{z_{1}} + 1 \right) g_{1}  \cdot f =   g_{2}^* g_{3}^*, \nonumber \\
&& \left( {\rm{i}} D_{z_{2}} + 1 \right) g_{2}  \cdot f =   g_{1}^* g_{3}^*, \label{HiorderBiline2}  \\
&& \left( {\rm{i}} D_{z_{3}} + 1 \right)g_{3}  \cdot f =    g_{1}^* g_{2}^*. \nonumber
\end{eqnarray}

Below, we will first construct algebraic solutions to the more general bilinear system
\begin{eqnarray}
&& \left( {\rm{i}} D_{z_{1}} + 1 \right) g_{1}  \cdot f =   h_2h_3, \nonumber \\
&& \left( {\rm{i}} D_{z_{2}} + 1 \right) g_{2}  \cdot f =   h_1h_3, \label{HiorderBiline2b}  \\
&& \left( {\rm{i}} D_{z_{3}} + 1 \right)g_{3}  \cdot f =    h_1h_2, \nonumber
\end{eqnarray}
where $h_1, h_2$ and $h_3$ are also complex functions. Afterwards, we will impose the reality conditon for $f$ and complex conjugation conditions
\[\label{NonlocPara}
h_{k}^*=g_{k}, \quad k=1, 2, 3.
\]
Under these conditions, the bilinear system (\ref{HiorderBiline2b}) will become the bilinear system (\ref{HiorderBiline2}), and the corresponding algebraic solutions will give rational solutions of the three-wave system (\ref{3WRIModel}) through the above variable and coordinate transformations (\ref{Transform}) and (\ref{VariaTrans1}).

Gram solutions to the general bilinear system (\ref{HiorderBiline2b}) have been studied in Ref.~\cite{YangYang3wave}. After simple generalizations of those results, we find that the $\tau$ function
\[\label{Mijdeterminants}
\tau_{n,k}=\det_{1\leq i,j \leq N} \left(m_{i,j}^{(n,k)}\right),
\]
where
\begin{eqnarray} \label{mij-diff2}
&& m_{i,j}^{(n,k)}=\left.
  \frac{\left(p\partial_{p}\right)^{n_{i}}}{ (n_{i}) !}\frac{\left(q \partial_{q}\right)^{n_{j}}}{(n_{j}) !}
  \widetilde{m}_{i,j}^{(n,k)}\right|_{p=p_{i}, \ q=q_{j}},
\end{eqnarray}
\begin{eqnarray}
&& \widetilde{m}_{i,j}^{(n,k)}=
  \frac{1}{p + q}\left(-\frac{p}{q}\right)^{k}\left(-\frac{p-\rm{i}}{q+\rm{i}}\right)^{n} \mathrm{e}^{\xi_i + \eta_j}, \\
&& \xi_i =\frac{1}{p} z_1 + \frac{1}{p-\rm{i}} z_2 + (p-{\rm{i}}) z_3 + \xi_{0,i} , \label{defxi} \\
&& \eta_j=\frac{1}{q} z_1 + \frac{1}{q+\rm{i}} z_2 + (q+{\rm{i}}) z_3 + \eta_{0,j},  \label{defeta}
\end{eqnarray}
$p_i, q_j$ are arbitrary complex constants, and $\xi_{0,i}(p)$, $\eta_{0,j}(q)$ are arbitrary functions of $p$ and $q$, respectively, would satisfy the following bilinear system
\begin{eqnarray}
&& \left[ \textrm{i} D_{z_1} + 1 \right] \tau_{n+1, k} \cdot \tau_{n, k} = \tau_{n, k+1} \tau_{n+1, k-1}, \nonumber \\
&&  \left[ \textrm{i} D_{z_2} + 1 \right] \tau_{n, k-1} \cdot \tau_{n, k} = \tau_{n-1, k} \tau_{n+1, k-1}, \label{HdimBilineEq2}\\
&& \left[  \textrm{i} D_{z_3} + 1  \right] \tau_{n-1, k+1} \cdot \tau_{n, k} = \tau_{n-1, k} \tau_{n, k+1}. \nonumber
\end{eqnarray}
Thus, if we define
\[ \label{fgdef1}
f=\tau_{0,0}, \quad g_{1}=\tau_{1,0}, \quad  g_{2}=\tau_{0,-1}, \quad g_{3}=\tau_{-1,1},
\]
and
\[ \label{fgdefhi}
h_{1}=\tau_{-1,0}, \quad  h_{2}=\tau_{0,1}, \quad h_{3}=\tau_{1,-1},
\]
the above bilinear system (\ref{HdimBilineEq2}) would reduce to (\ref{HiorderBiline2b}), and these $\tau$ functions would be solutions to that bilinear system.

To introduce explicit free parameters into the above solutions, we let
\[ \label{xieta}
\xi_{0,i}=\sum _{r=1}^\infty  a_{r,i} \ln^r \left[ \frac{p}{p_{i}} \right], \quad \eta_{0,j}=\sum _{r=1}^\infty b_{r,j}  \ln^r \left[ \frac{q}{q_{j}}\right],
\]
where $a_{r,i}$ and $b_{r,j}$ are free complex constants.

In order to reduce the bilinear system (\ref{HiorderBiline2b}) to the original one in (\ref{HiorderBiline2}), we still need to impose the $f$-reality condition as well as the complex conjugacy conditions of $h_{i}=g_{i}^*$. All these conditions would be satisfied if
\[ \label{condtaunk}
\tau_{n,k}= \left[ \tau_{-n,-k}\right]^*.
\]
To realize this condition, we set
\[ \label{para_cond}
q_j=p_j^*,     \quad b_{r,j}=a^*_{r,j}.
\]
In this case, we can readily show that
\[ \label{mijsym}
m_{j,i}^{(-n, -k)} = \left[m_{i,j}^{(n,k)}\right]^*.
\]
Thus, the condition (\ref{condtaunk}) holds, and the $\tau$ functions (\ref{fgdef1}) are solutions to the original bilinear system (\ref{HiorderBiline2}). Inserting the above $(\xi_{0,i}, \eta_{0,j})$ expressions (\ref{xieta}) and parameter conditions (\ref{para_cond}) into the matrix element expression (\ref{mij-diff2}), we then obtain the rational solutions in Theorem 1 for the three-wave system (\ref{3WRIModel}).

\subsection{Proof of Theorem 2}
Next, we remove differential operators in solutions of Theorem 1, and derive a more explicit solution form that is given in Theorem 2.

The technique we use is similar to that developed in \cite{OhtaJY2012,YangYang3wave}. We first introduce the generator $\mathcal{G}$ of differential operators $(p \partial_{p})^{i}  (q \partial_{q})^{j}$ as
\[
\mathcal{G}= \sum_{i=0}^\infty \sum_{j=0}^{\infty} \frac{\kappa^i}{i!} \frac{\lambda^j}{j!} \left( p\partial_{p} \right)^{i}  \left( q \partial_{q} \right)^{j}.   \nonumber
\]
For this generator, we have the identity \cite{OhtaJY2012}
\[
\mathcal{G} F(p, q)=F(e^{\kappa}p, e^{\lambda}q),   \nonumber
\]
where $F(p,q)$ is an arbitrary function. Then,
\begin{eqnarray*}
&& \frac{1}{(p+q)\widetilde{m}_{i,j}^{(n,k)}}\left. \mathcal{G} \widetilde{m}_{i,j}^{(n,k)}\right|_{p=p_{i},\  q=q_{j}}=\frac{1}{p_{i}  e^{\kappa}+ q_{j}  e^{\lambda}} \left(\frac{e^{\kappa}}{e^{\lambda}} \right)^k  \left( \frac{ p_{i} e^{\kappa}-\rm{i}}{p_{i}-\rm{i}} \right)^n \left( \frac{q_{j} e^{\lambda}+\rm{i}}{q_{j}+\rm{i}}  \right)^{-n} \exp \left(\sum _{r =1}^\infty (a_{r,i} \kappa^{r} +a^*_{r,j} \lambda^{r})\right) \times  \nonumber \\
&&  \hspace{1cm} \exp\left\{    \left( \frac{1}{p_{i} e^{\kappa}} - \frac{1}{p_{i}}+ \frac{1}{q_{j} e^{\lambda}}-\frac{1}{q_{j}} \right)z_1 + \left( \frac{1}{p_{i} e^{\kappa}-\rm{i}} - \frac{1}{p_{i}-\rm{i}}+ \frac{1}{q_{j} e^{\lambda}+\rm{i}}-\frac{1}{q_{j}+\rm{i}} \right)z_2+\left(p_{i} e^{\kappa}-p_{i}+ q_{j} e^{\lambda} -q_{j} \right)z_3  \right\}.   \hspace{1cm}
\end{eqnarray*}
The $1/(p_ie^\kappa+q_je^\lambda)$ term above can be rewritten as \cite{YangYang3wave}
\begin{eqnarray*}
&& \frac{1}{ p_i e^{\kappa} + q_j e^{\lambda} }=  \frac{p_{i}+q_j}{(p_{i} e^{\kappa}+q_j)(q_j e^{\lambda}+p_{i})} \sum_{\nu=0}^{\infty} \left[ \frac{(p_{i} e^{\kappa}-p_{i})(q_j e^{\lambda} -q_j)}{( p_{i} e^{\kappa}+q_j)(q_j e^{\lambda}+p_{i})} \right]^{\nu} \\
&& =  \frac{p_{i}+q_j}{(p_{i} e^{\kappa}+q_j)(q_j e^{\lambda} +p_{i})} \sum_{\nu=0}^{\infty} \left( \frac{p_{i}  q_j }{(p_{i} + q_j )^2} \kappa \lambda  \right)^{\nu}
\left( \frac{p_{i}  + q_j }{\kappa}  \frac{ e^{\kappa} -1}{p_{i} e^{\kappa}+q_j}  \right)^{\nu}
\left( \frac{p_{i} + q_j }{\lambda}  \frac{e^{\lambda}-1}{q_j e^{\lambda}+p_{i}}  \right)^{\nu} \\
&& =  \sum_{\nu=0}^{\infty} \left(\frac{1}{p_{i}+q_{j}}\right) \left( \frac{p_{i} q_j }{(p_{i}+q_j)^2} \kappa \lambda  \right)^{\nu}
\left(\frac{p_{i}+q_j}{p_{i} e^{\kappa}+q_j}\right) \left(\frac{p_{i}+q_j}{q_j e^{\lambda} +p_{i}}\right)
\left( \frac{p_{i}  + q_j }{\kappa}  \frac{ e^{\kappa} -1}{p_{i} e^{\kappa}+q_j}  \right)^{\nu}
\left( \frac{p_{i} + q_j }{\lambda}  \frac{e^{\lambda}-1}{q_j e^{\lambda}+p_{i}}  \right)^{\nu}.
\end{eqnarray*}
Using the above two equations as well as expansions (\ref{schucoefalpha})-(\ref{schurcoeffsr}), together with the conjugation relation $q_j=p_j^*$, we get
\begin{equation*}
\hspace{-1.5cm}\frac{1}{(p+q)\widetilde{m}_{i,j}^{(n,k)}}\left. \mathcal{G} \widetilde{m}_{i,j}^{(n,k)}\right|_{p=p_{i},\  q=q_{j}}=\sum_{\nu=0}^{\infty} \left(\frac{1}{p_{i}+q_{j}}\right) \left( \frac{p_{i} q_j }{(p_{i}+q_j)^2} \kappa \lambda  \right)^{\nu} \exp\left(
\sum_{r=1}^\infty \left(x^+_{r,i,j}(n,k)+\nu s_{r, i, j}\right)\kappa^r +
\sum_{r=1}^\infty \left(x^{-}_{r,j,i}(n,k)+\nu s^*_{r, j, i}\right)\lambda^r\right),
\end{equation*}
where $x^\pm_{r,i,j}(n,k)$ and $s_{r, i, j}$ are as defined in Theorem 2. Taking the coefficients of $\kappa^i\lambda^j$ on both sides of the above equation, we get
\begin{equation} \label{mijscaling}
\frac{m_{i,j}^{(n, k)}}{(p_i+p_j^*)\widetilde{m}_{i,j}^{(n,k)}}=\sum_{\nu=0}^{\min(n_{i}, n_{j})}\left(\frac{1}{p_{i}+p_{j}^*}\right) \left[ \frac{ p_{i} p_{j}^* }{(p_{i}+p_{j}^*)^2}  \right]^{\nu} \hspace{0.06cm} S_{n_{i}-\nu}[\textbf{\emph{x}}^{+}_{i,j}(n,k) +\nu \textbf{\emph{s}}_{i,j}] \hspace{0.06cm} S_{n_{j}-\nu}[\textbf{\emph{x}}^{-}_{j,i}(n,k) + \nu \textbf{\emph{s}}^*_{j,i}],
\end{equation}
where $m_{i,j}^{(n,k)}$ is the matrix element defined in Eq. (\ref{mij-diff2}). It is easy to see that the function $\tau_{n,k}$, with its matrix element $m_{i,j}^{(n,k)}$ scaled as the left side of the above equation, still satisfies the bilinear equations (\ref{HiorderBiline2}), because the scaling factor is an exponential of a linear function in $(z_1, z_2, z_3)$. Thus, the scaled $\tau_{n,k}$ function, with its matrix elements given by the right side of the above equation, i.e., by Eq.~(\ref{Schmatrimnij}) of Theorem 2, still satisfies the bilinear equations (\ref{HiorderBiline2}) of the three-wave system. This completes the proof of Theorem 2.

\subsection{Proof of Theorem 3}
To derive semi-rational solutions in the three-wave system (\ref{3WRIModel}), we choose the matrix element $m_{i,j}^{(n,k)}$ of the $\tau_{n,k}$ function as in Eq.~(\ref{mij-diff5}), which is the same as that in Eq.~(\ref{mij-diff2}) plus a complex constant $\pi_{i,j}$. It is easy to see that such a $\tau_{n,k}$ function still satisfies the bilinear system (\ref{HdimBilineEq2}) for reasons explained in Ref.~\cite{YangYang3wave}. Under the condition $\pi_{i,j}=\pi^*_{j,i}$, the corresponding $\tau_{n,k}$ function satisfies the complex-conjugation condition (\ref{condtaunk}) as well. Thus, this $\tau_{n,k}$ function with the matrix element $m_{i,j}^{(n,k)}$ as given in Eq.~(\ref{mij-diff5}) satisfies the bilinear equations (\ref{HiorderBiline2}) of the three-wave system. To derive the more explicit expression (\ref{Schmatrimncij}) of $m_{i,j}^{(n,k)}$ in Theorem 3, we just need to repeat the calculations in the proof of Theorem 2, except that the $\pi_{i,j}$ term in $m_{i,j}^{(n, k)}$, after the scaling of $(p_i+p_j^*)\widetilde{m}_{i,j}^{(n,k)}$ in Eq.~(\ref{mijscaling}), gives rise to the exponential terms in Eq.~(\ref{Schmatrimncij}) of Theorem 3. This theorem is then proved.

\section{Summary and Discussion}

In this paper, we have investigated rogue waves in the (2+1)-dimensional three-wave system (\ref{3WRIModel}) by the bilinear method. General rogue waves arising from a constant background, from a lump-soliton background and from a dark-soliton background have been derived, and their dynamics illustrated. We have shown that for rogue waves arising from a constant background, fundamental rogue waves are line-shaped, and multi-rogue waves exhibit multiple intersecting lines. In addition, higher-order rogue waves could also be line-shaped, but would exhibit multiple parallel lines. For rogue waves arising from a lump-soliton background, they could exhibit distinctive patterns due to their interaction with the lump soliton. For rogue waves arising from a dark-soliton background, their intensity pattern could feature half-line shapes or lump shapes, which are very novel.

In this article, we have only illustrated the simplest rogue wave solutions. It is known that as the order $N$ increases, rogue waves could exhibit more striking patterns \cite{YangYang2021b}. Whether rogue waves in the (2+1)-dimensional three-wave system (\ref{3WRIModel}) also display novel patterns at large $N$ values is an interesting question which merits further study.

\section*{Acknowledgment}
The work of J.Y. was supported in part by the National Science Foundation under award number DMS-1616122.

\end{document}